\documentclass[openbib,12pt]{article}
\input{psfig.sty}
\usepackage{amssymb,amsmath,epsfig}
\def\doublespace{\parskip 3pt plus 1.2pt
     \baselineskip 18pt plus 1pt minus .5pt
     \lineskip 2pt plus 1pt \lineskiplimit 5pt}

\begin{document}

                              
\begin{center} 
{\Large {\bf Topology of Central Pattern Generators:\\
Selection by Chaotic Neurons}}\\

\end{center} 
\begin{center}

\bigskip
\bigskip
\bigskip
{ R. Huerta, P. Varona}\\
\bigskip
Institute for Nonlinear Science\\
E.T.S. de Ingenier\'{\i}a Inform\'{a}tica,\\
Universidad Aut\'{o}noma de Madrid, 28049 Madrid (SPAIN).\\
huerta@routh.ucsd.edu, pvarona@lyapunov.ucsd.edu\\
\bigskip
\bigskip
{ M. I. Rabinovich}\\
\bigskip
Institute for Nonlinear Science\\
rabin@landau.ucsd.edu\\
\bigskip
and\\
\bigskip
{ Henry D. I. Abarbanel\footnote{Institute for Nonlinear Science}}\\
\bigskip
Department of Physics and Marine Physical Laboratory,\\
Scripps Institution of Oceanography,\\
hdia@jacobi.ucsd.edu
\bigskip

\bigskip
\bigskip
\bigskip
University of California, San Diego\\
La Jolla CA 92093-0402 USA\\

\end{center}
\clearpage
\doublespace
 \begin{center}
{\large {\bf Abstract}} 
\end{center}

Central Pattern Generators (CPGs) in invertebrates are comprised 
of networks of neurons in which every neuron has reciprocal connections to 
other members of the CPG. This is a ``closed'' network topology. 
An ``open'' topology, where one or more neurons receives input but 
does not send output to other member neurons, is not found in these CPGs. 
In this paper we investigate a possible reason for this topological
structure using the ability to perform a biological functional task as
a measure of the efficacy of the network. When the CPG is composed of
model neurons which exhibit regular membrane voltage oscillations,
open topologies are essentially as able to maximize this functionality
as closed topologies. When we replace these models by neurons which
exhibit chaotic membrane voltage oscillations, the functional
criterion selects closed topologies when the demands of the task are
increased, and these are the topologies observed in known CPG
networks. As isolated neurons from invertebrate CPGs are known in some
cases to undergo chaotic oscillations~\cite{HI,aba}
 this provides a biological basis for 
understanding the class of closed network topologies we observe.

\clearpage

\section{Introduction}

Central Pattern Generator (CPG) neural networks in invertebrates perform 
a wide variety of functional roles, each of them requiring rhythmic 
output from the CPG to the muscles controlling the function \cite{Marder}. 
In our study of the pyloric CPG of the California spiny lobster 
{\em Panrulis interruptus} we have addressed the question why the
network is closed in the sense that all neurons in the CPG both send 
signals to other neurons in the circuit and receive input from
other neurons in the circuit. None of the neurons only receive
from or send signals to the rest of the network. The pyloric CPG
is illustrated in the right half of Figure 1A.
\vspace*{0.1cm}
\begin{figure}[ht!]
\centerline{
\epsfbox{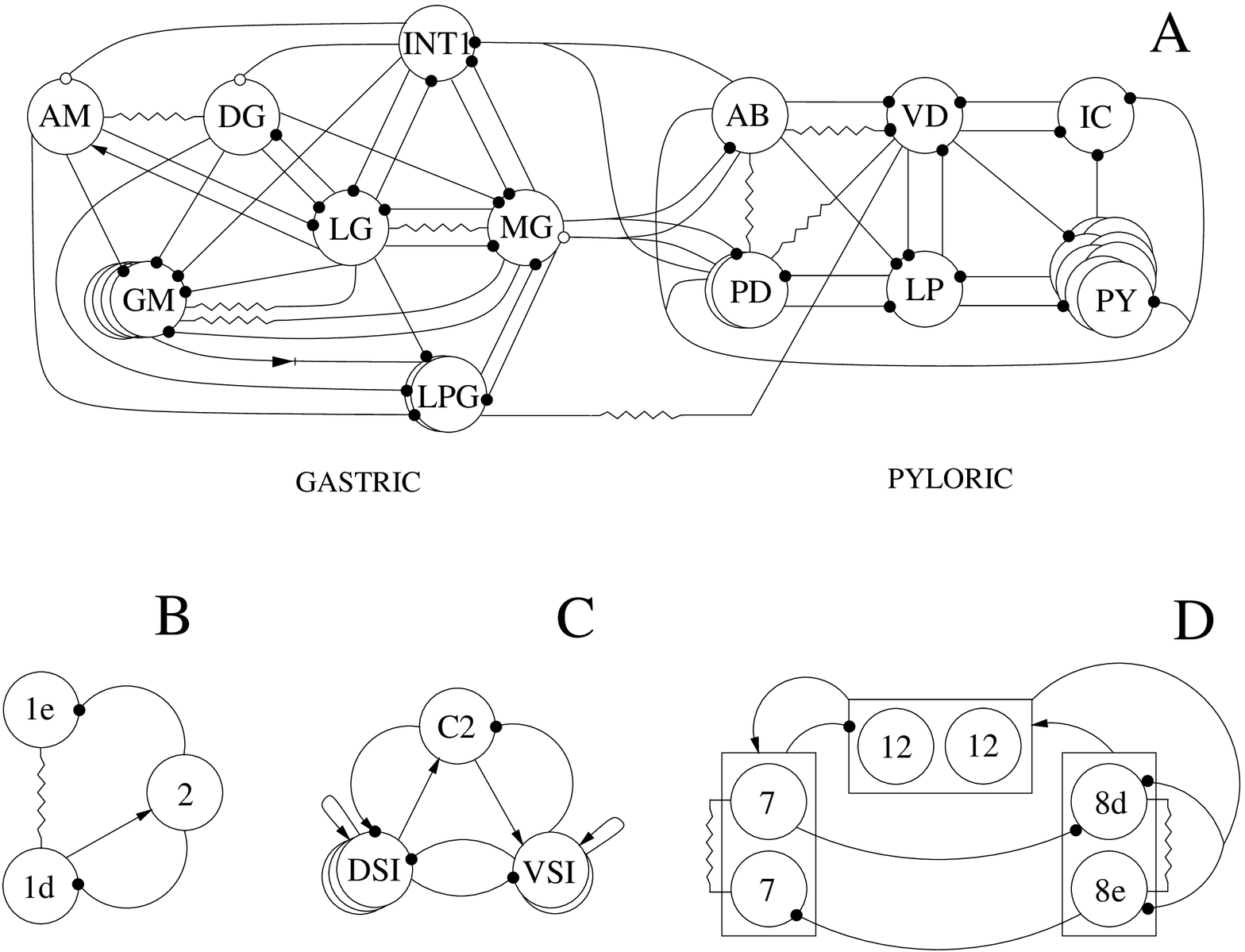}
}
\caption{\footnotesize Examples of closed topologies in invertebrate
CPGs: (A) the gastric and pyloric CPGs in crustacea (modified
from~\cite{selverston}), (B) The feeding CPG in {\sl Planorbis}
(modified from \cite{ara}), (C) The swimming CPG in {\sl Tritonia}
(modified from \cite{Getting}), (D) The swimming CPG in {\sl Clione}
(modified from \cite{arb}. Dots represent inhibitory synapses
while arrows represent excitatory synapses. Electrotonic gap junctions
are represented by resistors.}
\label{cpgs}
\end{figure}
\vspace*{2mm}

In this paper we address the appearance of this closed CPG network and 
seek a functional biological reason why evolution may have been led to 
select this configuration.
We are aware of investigations \cite{Getting} which attempt to answer 
this question in terms of enhancing robustness or self-organization. 
In the present work we test an alternate hypothesis: {\em the closed
network topology composed of irregular neurons provides greater
efficiency in the task which the network  is required to perform}.  We present here evidence for this hypothesis in terms of calculations of the ability of a model crustacean pyloric CPG to perform its task of transporting shredded food from the stomach to the digestive system. We build the CPG first from conductance based Hodgkin-Huxley (HH)~\cite{HH,huerta1} neurons which exhibit periodic oscillations and subsequently from model neurons of Hindmarsh-Rose (HR)~\cite{HR} type which have chaotic membrane voltage oscillations. Using the transport criterion made explicit below, we find the periodic HH neurons to allow open topologies in CPG operation while chaotic component neurons allow only closed topologies on our functional criterion.

In an earlier paper~\cite{huerta2} we described in some detail the
use of regular HH neurons with two spatial compartments as CPG
network components. Many of the allowed network configurations,
were open and quite robust in the presence of noise as the HH models oscillated in a strongly dissipative limit cycle and did not sit in parameter space near a bifurcation point. In real biological networks the component neurons show substantial irregularity in their bursting cycles~\cite{elson1,elson2}, and this has led us to the work reported here.

The chaotic model neuron we use in our investigations is a three
dimensional version of an HR neuron. It exhibits spiking-bursting behavior as 
observed in the laboratory, and this alone would not distinguish it 
from similar behavior of HH type neuron models. Some evidence, peripheral 
to this study, for the realistic structure of the HR neurons comes from 
its observed ability to act as a substitute for biological neurons in 
the lobster pyloric CPG when it is realized in simple analog circuitry and coupled into that biological network~\cite{szucs}. The main feature of the chaotic
neuron is that it is built around a homoclinic loop, which carries the fast oscillations. Motion on the homoclinic loop is intrinsically unstable which means that any perturbation close to the homoclinic structure will produce striking differences in its membrane voltage time course: the neuron will either fire another spike or hyperpolarize the 
membrane~\cite{baz,aba,wang}. The dynamical pictures of the 
chaotic and the regular models are completely different.

The functional criterion we investigate is based on the role of the
pyloric CPG in the California spiny lobster. The pyloric CPG has as
its task the transport of shredded food from the stomach to the
digestive system. Our functional criterion for the selection of
network topologies is the ability to perform this transport as the
severity of the task is increased. In an intuitive way one might
expect to maximize this transport through regular CPG oscillations,
and this is consistent with a fully coupled or closed topology of
interconnections which regularize the chaotic behavior of individual
components of the CPG assembly~\cite{baz}. Our overall picture then suggests a correspondence between the functional task of a CPG, the observations of regular patterned behavior in CPG assemblies of individually chaotic components, and the CPG network topology.

\section{The chaotic model neuron}

In our simulations we used a modified, three dimensional version of the HR model neurons. The model is comprised three dynamical variables
comprising a fast subset, $x(t)$ and $y(t)$, and a slower $z(t)$. $x(t)$ represents the cell's membrane potential. These dynamical variables satisfy

\begin{eqnarray}
\frac{dx(t)}{dt} & = &
4y(t) + 1.5x^{2}(t)-0.25x^{3}(t)-2z(t)+2e+I_{\mbox{syn}}\\
\frac{dy(t)}{dt} & =
& 0.5- 0.625x^{2}(t)-y(t),\\
\frac{1}{\mu}\frac{dz(t)}{dt} & = &
- z(t)+2  [ x(t)+3.2] 
\label{eq1}
\end{eqnarray}
where $e$ represents an injected DC current, and
$\mu$ is the parameter that controls the time constant of the slow
variable. The parameters were chosen to place the isolated neurons in
the chaotic spiking-bursting regime: $e=3.281$,
$\mu=0.0021$. 

$I_{syn}$ represents the postsynaptic current evoked
after the stimulation of a chemical graded synapse. In this paper we
consider only inhibitory synapses, which is the main kind of
interconnection present in the pyloric CPG of the lobster. The
synaptic current has been simulated with the traditional description
used in the dynamical clamp technique~\cite{sharp} with minor
modifications:
\begin{equation}
I_{\mbox{syn}}  =  - g r(x_{\mbox{pre}}) \vartheta (x_{\mbox{post}})
\label{eq2}
\end{equation}
where $g$ is the maximal synaptic conductance, and $x_{\mbox{post}}$ is the membrane potential of the postsynaptic neuron.
$r(x_{pre})$ is the synaptic activation variable determined
from the presynaptic activity by:
\begin{eqnarray}
\frac{dr}{dt} & = & [r_{\infty}(x_{pre})-r]/\tau_{r} \\
r_{\infty}(x_{pre}) & = & [1+\tanh((x_{pre}+A)/B)]/2.
\label{eq3}
\end{eqnarray}
In our work $A = 1,2$ and $B = 0.9$.
$\tau_{r}$ is the characteristic time constant of the
synapse ($\tau_{r} \sim 100$). 

$\vartheta (x_{post})$ is a
nonlinear function of the membrane potential of the postsynaptic
neuron:
\begin{eqnarray}
\vartheta(x_{post}) & = & (1+\tanh((x_{post}+a)/b)). \\
\label{eq4}
\end{eqnarray}
The parameters for $\vartheta (x_{post})$ ($a=2.807514$ and $b=0.4$ in this paper) were
chosen so that this function remained linear for the slow
oscillations, the subthreshold regions for the fast spikes:

\begin{eqnarray}
\frac{d\vartheta(x)}{dx} \bigg\vert_{a,b} & \sim & 1,\\
\vartheta(x) \bigg\vert_{a,b} & \sim & 0.
\label{eq5}
\end{eqnarray}

\begin{figure}[ht!]
\centerline{\epsfxsize=2.5in
\epsfbox{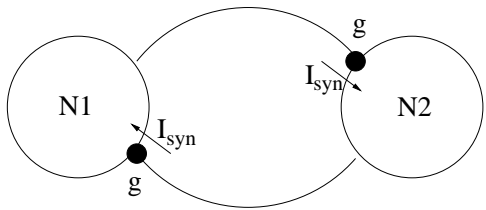}
}
\caption{\footnotesize Basic configuration of synaptic couplings of the model for
mutual inhibition between two neurons. The graded synapses are described by
equations (\ref{eq2}-\ref{eq4}) in the text.}
\label{2synapses}
\end{figure}

Each pair of model neurons were connected in mutual inhibitory configurations
consisting of basic two-cell units such as the one shown in
Figure~\ref{2synapses}.

\vspace*{0.1cm}

\section{The models for the CPG and the mechanical device}

We have previously built a CPG that controls a pyloric chamber model
~\cite{huerta2}. The CPG is composed of three neurons with mutual
inhibitory coupling as shown in Fig. 2. The configuration we selected is shown in Figure 3. The six independent maximal conductances $g_{ij}; \;i\ne j;\; i,j=1,2,3$ are to be selected 
as described below. The CPG sends electrical
activity to the `muscles' which control the dilation and contraction
of the `pyloric chamber' represented here by a simulated mechanical
`plant'. Three neurons are the minimal number of neurons that produces 
a maximization of the average flow of food pumped out of the plant.

\begin{figure}[ht!]
\centerline{\psfig{figure=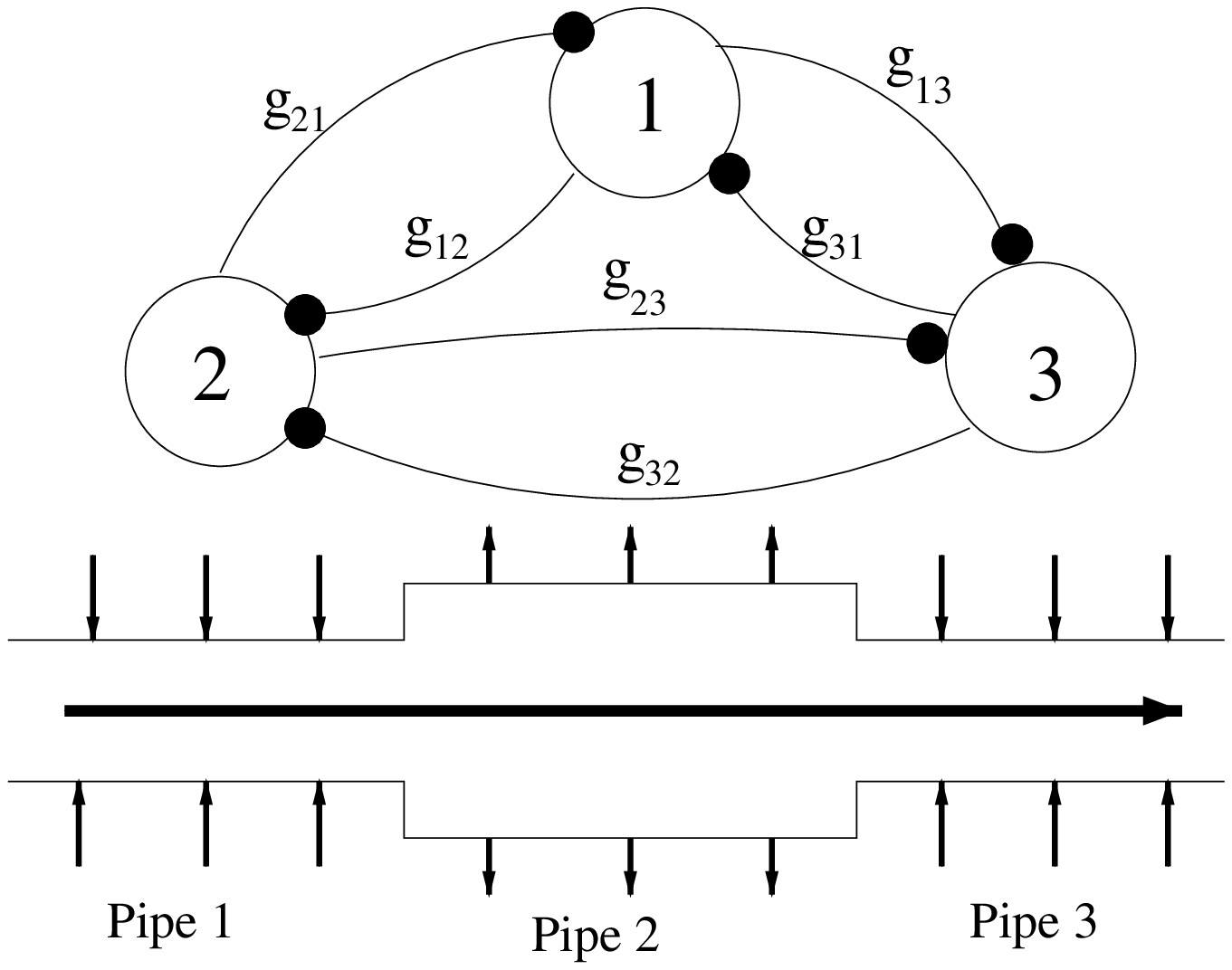,width=10cm}}
\caption{Illustration of the mechanical plant and the CPG. In the upper part of this figure we show the three neuron CPG and the six possible connections $g_{ij}$ among them using mutual inhibition. In the lower half of the figure we show the pumping `plant' composed of three pipes which scan vary their radius.  Each CPG neuron only
affects one segment of the pipe. A detailed description of the pump is 
given in~\cite{ huerta2}. 
\label{fig1}}
\end{figure}

The mechanical plant is inspired by the lobster pyloric 
chamber~\cite{selverston}. In Fig. 3 we have sketched the model
with three joined pipes that can
expand or contract radially. The central pipe 
only dilates and the end pipes only contract; this mimics the behavior of
the pyloric chamber of the lobster.
The proper dynamical combination of muscle activity influencing the walls of the pipes will lead to movement of shredded food towards the right end of the pyloric chamber.

In our selection procedure for combinations of the $g_{ij}$, we set the average transport through the pump in time T to a set of increasingly larger values and determine which configurations are able to achieve each level of transport. 
The average throughput over time $T$ is given by
\begin{equation}
\Phi(g_{ij}) = \frac{\rho}{T}\int_{t_0}^{t_0+T} A_3(t)  v_6(t)dt,
\label{eq:1}
\end{equation}    
where  $\rho$ is the density of the material being pumped, $A_3(t)$ is the cross sectional area of the rightmost pipe section, $v_6(t)$ is the mean velocity through the rightmost cross section of the plant, and $t_0$ is some starting time. The differential equations determining $A_3(t)$ and $v_6(t)$ were derived in~\cite{huerta2}. 
They come from the Navier-Stokes equations and from mass and energy conservation.
We treat the shredded food as homogeneous,
incompressible and isothermal, and it moves with low Reynolds number so the flow is always laminar and all radial velocities in the pipe sections are small.  We also assume that no food leaks out during the pumping and that there were no
head-losses in the joints of the pipes. 

In our calculations the mean initial velocities in the sections are zero. We choose a set of $g_{ij}$ as indicated in a moment, and ask whether the specified configuration can achieve a chosen level of $\Phi(g_{ij})$. The system of differential equations was integrated using a Runge-Kutta 6(5) scheme with variable time step and with an absolute error of $10^{-16}$ and a relative error of $10^{-6}$.

We choose the $g_{ij}$ to be 0, 50nS, or 200nS, and evaluate $\Phi(g_{ij})$ 
for each of these $3^6$ choices.  The system was run for a simulated time 
of 60 sec; a transient of $t_0 =\,$10 sec was eliminated and 
$\Phi(g_{ij})$ was found by averaging over the last $T =\,$50 sec.

In the following section we will compare the performance of both
regular and chaotic neurons in controlling the pump for the different
connection architectures. The detailed
description of the regular HH model can be found
in~\cite{huerta2}. The chaotic neurons were implemented as described
in the previous section. Both types of neurons are able to generate
spiking-bursting behavior.

\section{Analysis of solutions}
\begin{figure}[ht!]
\centerline{\psfig{figure=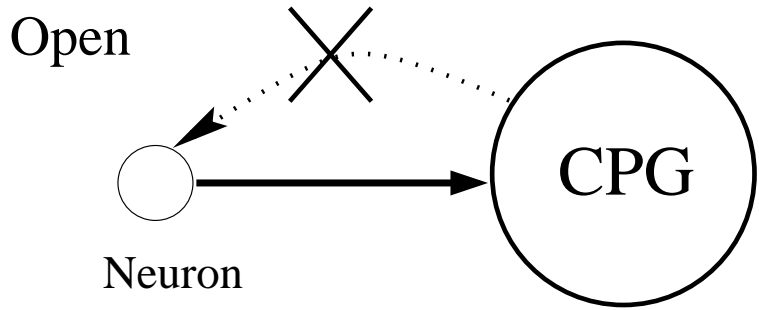,width=10cm}}
\caption{Definition of  an {\bf open} network topology:
there must be one or more neuron without feedback from the remaining neurons.
\label{fig2}}
\end{figure}
It is useful to distinguish among three different types of topologies: 

\begin{itemize}
\item {\bf Open topology}: in this case there is one or more neuron which 
receives no feedback from the other network elements.
This is shown in Figure 4. 
\item {\bf Semi-open topology}: in this case there is
one or more neuron which receives input from some neurons in the 
network, but no connection is made back to the network.
\item {\bf Closed topology}: in this case there are 
connections from each network member to and from the remainder of the network. 
\end{itemize}

In order to quantify the difference between open topologies 
and the others we introduce the following definitions. We define 
the set of solutions with a resulting flow lying in $\Phi_{lower} \le \Phi \le \Phi_{upper}$ as 
\begin{equation}
G(\Phi_{lower},\Phi_{upper})=\left\{(g_{12},g_{13},g_{21},g_{23},g_{31},g_{32});\Phi_{lower}<\Phi\leq \Phi_{upper} \right\}.
\end{equation}
and the subset of open solutions as 
\begin{equation}
O(\Phi_{lower},\Phi_{upper})=\left\{(g_{12},g_{13},g_{21},g_{23},g_{31},g_{32});\exists
i / g_{ji}=0, \forall j\neq i;\Phi_{lower}<\Phi\leq \Phi_{upper} \right\}.
\end{equation}
We report the ratio between the number of open solutions 
$\mid
O(\Phi_{lower},\Phi_{upper})\mid$ and the total number of solutions
$\mid G(\Phi_{lower},\Phi_{upper})\mid$ 

\begin{equation}
\eta(\Phi_{lower},\Phi_{upper})=
\begin{cases}
\mid O(\Phi_{lower},\Phi_{upper}) \mid/\mid
G(\Phi_{lower},\Phi_{upper})\mid & \text{if $\mid
G(\Phi_{lower},\Phi_{upper}) \mid \neq 0$}\\
0 &\text{if $\mid
G(\Phi_{lower},\Phi_{upper}) \mid = 0$}\\
\end{cases}
\label{eqn3}
\end{equation} as a quantitative measure of the topology of allowed solutions at a given value of $\Phi$. $\mid \ldots \mid$ denotes the cardinality of a set.
Since $\mid O(\Phi_{lower},\Phi_{upper}) \mid \leq \mid
G(\Phi_{lower},\Phi_{upper})\mid$,$0 \le \eta \le 1$.

We investigated two questions in the calculations reported here:
\begin{itemize}
\item Does the use of chaotic component CPG neurons filter out open solutions ?
\item Can noise eliminate open solutions when the component CPG neurons are regular oscillators?
\end{itemize}

\begin{figure}[ht!]
\centerline{\psfig{figure=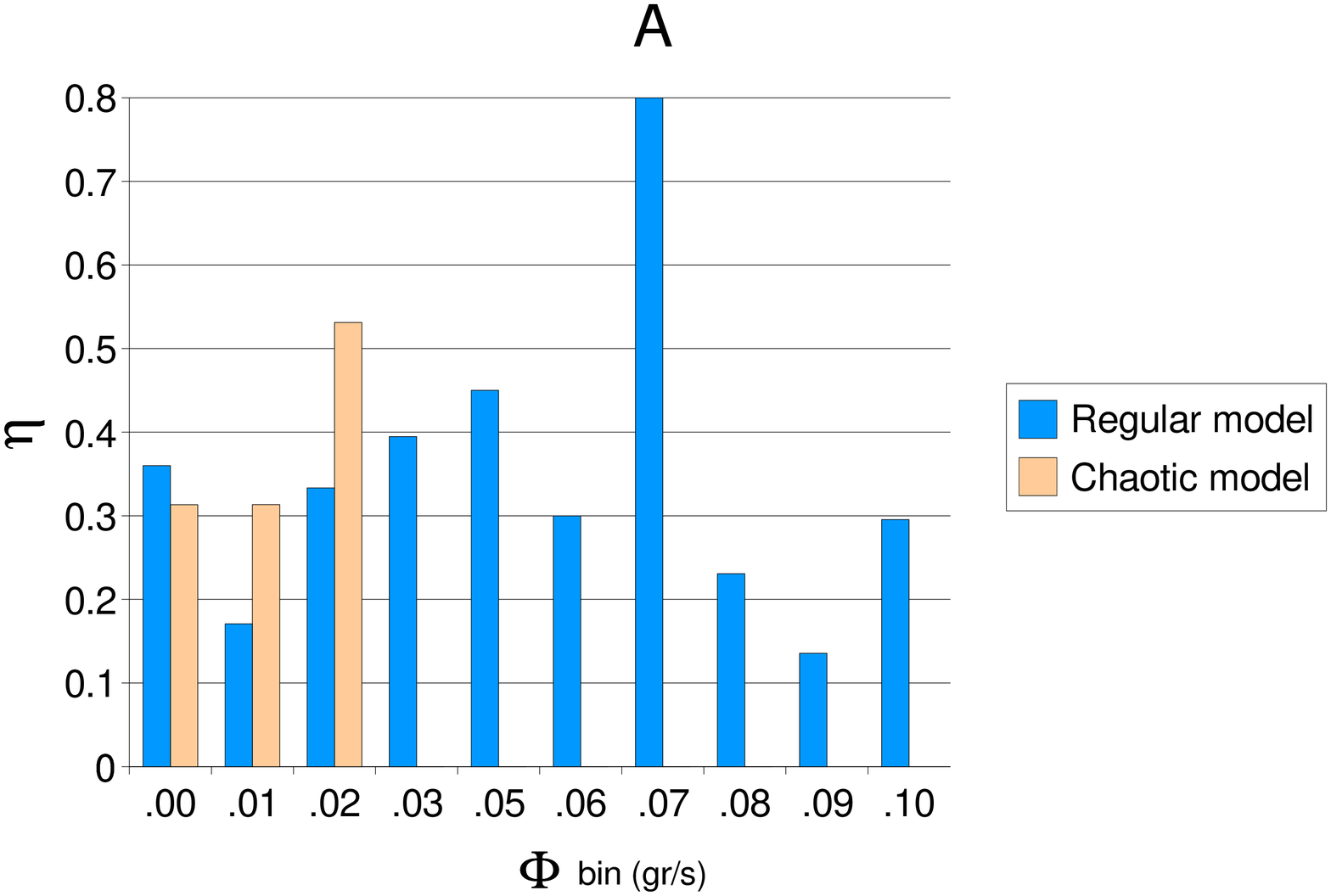,width=5in}}
\centerline{\psfig{figure=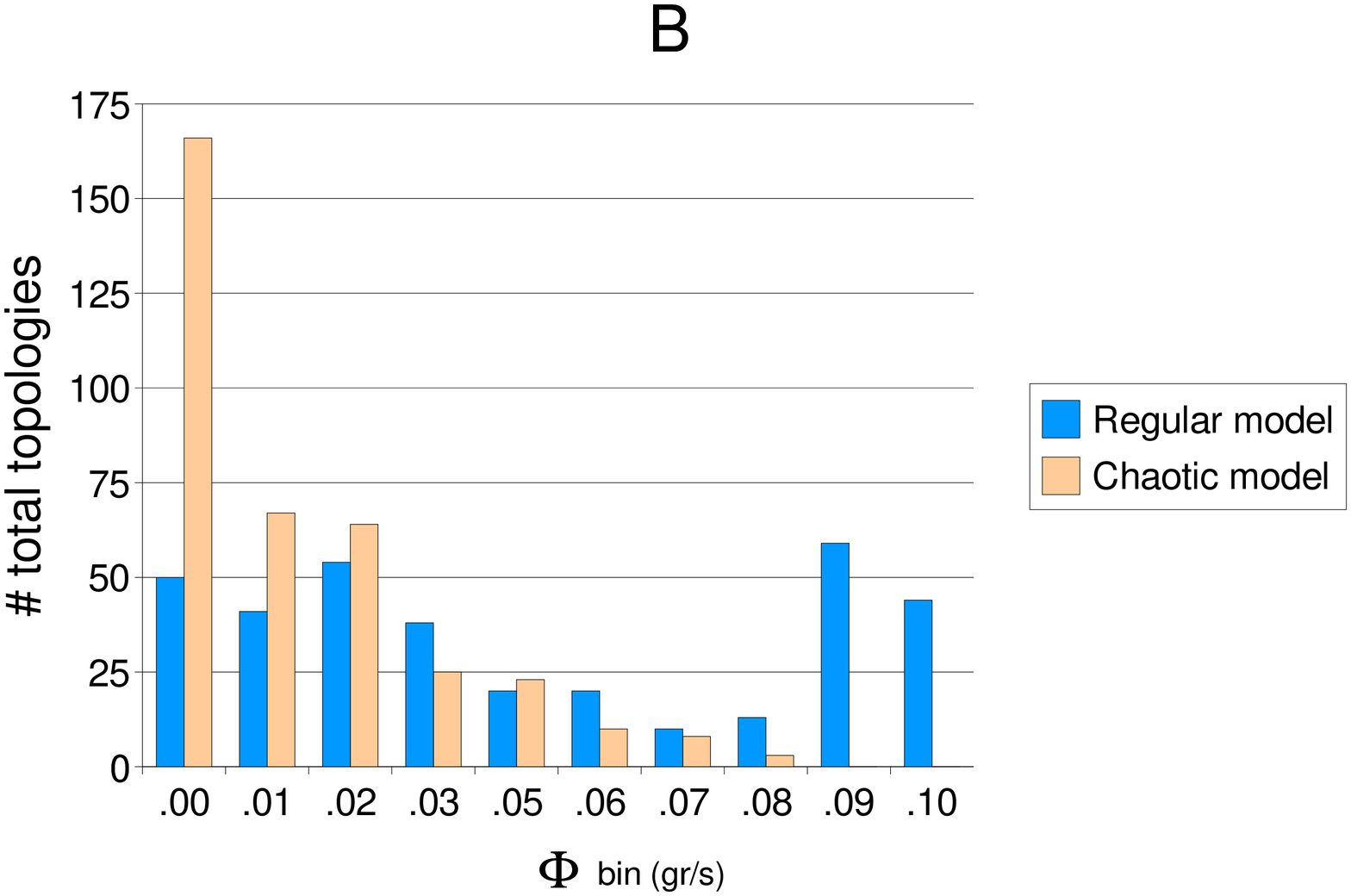,width=5in}}
\caption{{\bf A}. $\eta(\Phi_{\mbox{lower}},\Phi_{\mbox{upper}})$
for the regular and chaotic component CPGs. The data are presented by placing the average transport  in bins defined by $\Phi_{\mbox{lower}} = \Phi$ and $\Phi_{\mbox{upper}} = 1.01 \Phi$. As we increase the transport level $\Phi$, the number of open solutions which can achieve this level goes rapidly to zero. 
{\bf B}. Number of total allowed configurations for both the regular and chaotic CPG as a function of the level of average transport $\Phi$. 
 \label{fig6}}
\end{figure}

In examining the first question we used three different values of the $g_{ij}$ as indicated above for both the regular neuron model and the chaotic neuron model. The
number of configurations was $3^6$. The number of
simulation trials for a given configuration pattern was $15$ for
different random initial conditions.  In Figure \ref{fig6} we can see
a plot of $\eta(\Phi_{\mbox{lower}},\Phi_{\mbox{upper}})$
as defined in Equation~\ref{eqn3}.  We can see that as the flow $\Phi$ increases there are no open topologies of chaotic component neurons which can achieve the flow level.  $\Phi$ must be substantially reduced to find allowed open topology synaptic connection configurations. 

If we build the CPG with regular neurons we see that there are many open topology solutions for high values of $\Phi$. Moreover, the $\eta(\Phi)$ is strongly decreasing for the
CPG with chaotic neurons. However, the intrinsically regular CPG does
not lead to a strong dependence of $\eta$ on the level of the flow. 

This is one of the essential points of our work. Since the neurons in
the pyloric CPG appear to oscillate chaotically, a closed topology of
synaptic connections is required to achieve high transport rates for
food. Regular neurons can achieve high transport rates in open and in
closed configurations, but the pyloric  CPG is observed to be closed. 
Our calculations provide a suggestive connection between these
observations. CPGs made of chaotic neurons could have evolved open
topologies, but these are not as effective in transporting food as
closed topologies and presumably were not selected in evolutionary
processes. This means that by means of chaotic neurons a  selective
choice of available solutions is obtained.  Intrinsic chaotic dynamics in
the isolated neurons provide a straightforward explanation of the 
existence of non-open connection topologies in invertebrate's CPGs.

To address our second question iid N(0,1) white noise
$\epsilon(t)$ is introduced in the external injected currents of the Hodgkin-Huxley
model~\cite{huerta2} so $I(t)=I_o+\sigma \epsilon(t)$ where
$\sigma$ is the amplitude of the noise. We introduce this type 
of additive noise for simplicity, because it
can be more easily integrated~\cite{Mannella} since there is no
dependence on the other variables of the ordinary differential
equations.  In Figure 5 we can see the results of our calculations for $8$ different values of $\sigma$. For each $\sigma$ value 
20 trials of different initial conditions for a given configuration of $g_{ij}$
 were carried out and the average of those values was taken as the 
output of the calculation.
The quantity we used to estimate
changes as a function of $\sigma$ is 
\begin{equation}
\eta(\sigma)=
\begin{cases}
\mid O_{\sigma}(\Phi_{th},\Phi_{max}) \mid/\mid
G_{\sigma}(\Phi_{th},\Phi_{max})\mid,& \text{if $\mid
G_{\sigma}(\Phi_{th},\Phi_{max})\mid \neq 0$}\\
0 & \text{if $\mid
G_{\sigma}(\Phi_{th},\Phi_{max})\mid = 0$}\\
\end{cases}
\nonumber
\end{equation}
where $\Phi_{max}=\max_{{\bf g}}\{\Phi\}$ and $\Phi_{th}$ is the
threshold value that determines the existence 
of good solutions. If a reduction in the number 
of open configurations is observed the value $\eta(\sigma)$ must
decrease. If our  selected configurations are eliminated by noise, then
open-topology configurations should disappear and $\eta(\sigma)$
must tend to $0$. Otherwise for any value of noise then
$\eta(\sigma)$ is always greater than $0$. In fig 5 we can clearly see 
that for any value 
of the noise no reduction of the open-topology configurations is 
observed. Moreover, the quantity $\eta$ approaches $1$ for high 
values of the noise which answers our question about noise in the positive: the open configurations of regular component CPGs are robust against the presence of synaptic noise. The appearance of closed configurations must have a different explanation, and that we have provided above. 
\begin{figure}[ht!]
\centerline{\psfig{figure=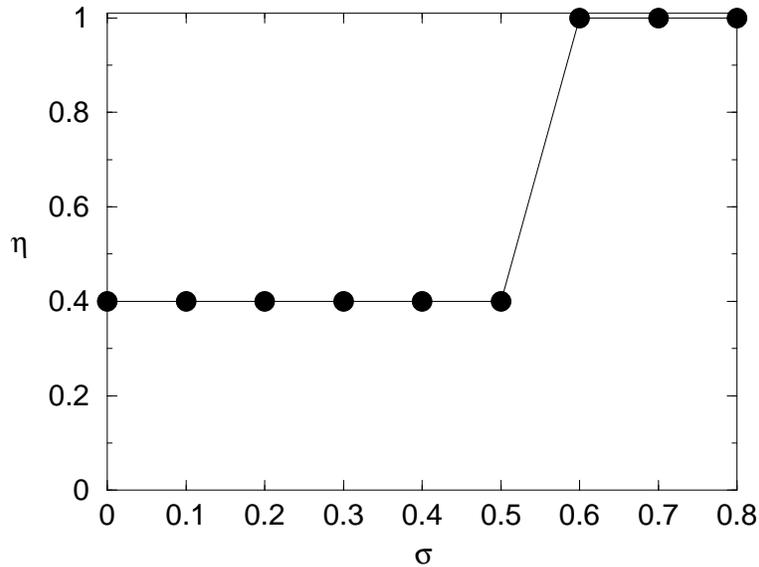,width=10cm}}
\caption{$\eta(\sigma)$ for regular Hodgkin-Huxley neurons as CPG components.
model. The open-topology configurations are not eliminated because
of the noise. In these calculations $\Phi=0.06$ (gr/sec). 
\label{fig5}}
\end{figure}

\section{Conclusion}

In this paper we have concentrated on the lobster pyloric CPG as
we have experimental evidence of the chaotic oscillations of its
component neurons when they are isolated from the intact network.
There are many other invertebrate CPGs as shown in Figure 1, and
a common thread among these CPGs, whether they are involved in shredding food, 
as in the gastric CPG of crustacea, feeding, as in the {\em Planorbis}, 
or swimming as in {\em Tritonia} is that the topology of the 
CPG network is  closed. It would be very interesting to follow
the path we have set out here for these and other CPGs to
establish whether the connection between network topology ,
efficiency in biological function and chaotic neuronal elements
holds there as well.

Since the early papers that showed evidence of chaos 
in CPG neurons~\cite{HI,Mpitsos} little attention has been given to the role of chaotic neurons in CPGs. It appears that the main reason is that intact CPG networks work in a regular fashion. Indeed their task is to provide a regular
rhythm to the animal so that the appropriate response of the muscles that operate in a physical device, such as the pyloric chamber, performs a specific function. Therefore, the interest of modelers was focused mainly on studying the dynamics of periodic model neurons~\cite{Skinner,Kepler,Terman1,Rinzel,Roberts,huerta1}. In very few cases
modelers have tried to understand the potential advantages of using chaotic neurons~\cite{Freeman,Rabin,Zak}, but typically in more complex systems, not in CPGs. By means of this work we hope to draw the reader's attention to an interesting set of roles played by chaotic neurons in CPGs. 

The connectivity of neurons in CPGs is probably determined by many developmental and evolutionary factors, and in this paper we have considered an aspect of the required function of a CPG as providing a significant driving force in this topological decision. Starting from the observation that component neurons from the lobster pyloric CPG have chaotic membrane potential oscillations when observed in isolation~\cite{aba}, we have investigated the influence that chaotic versus regular neurons might have on the structure of the neural interconnections in the CPG in order to achieve its functional goal. If our conclusions are to extend to other CPGs, clearly one must systematically, as we have for the pyloric CPG of lobster, establish the modes of oscillation for the CPG members in isolation. We conjecture that the example of lobster pyloric CPG is not special, but we are seeing in this example a pattern which will repeat in other CPG systems.
\clearpage

\section*{Acknowledgements} 
This work was supported in
part by the U.S. Department of Energy, Office of Basic Energy
Sciences, Division of Engineering and Geosciences, under grants
DE-FG03-90ER14138 and DE-FG03-96ER14592. We thank Rob Elson, Allen Selverston, and Attila Sz\"ucs for many conversations about the material in this paper.

\end{document}